# Astrometrical Observations of Pluto – Charon System with the Automated Telescopes of Pulkovo Observatory

*A.V. Devyatkin, E.A. Bashakova, V.Yu. Slesarenko*

Pulkovo observatory of RAS, Saint-Petersburg, Russia

The space probe "New Horizon" was launched on 19$^{th}$ of January 2006 in order to study Pluto and its moons. Spacecraft will fly by Pluto as close as 12500 km in the middle of July 2015 and will get the most detailed images of Pluto and its moon until this moment. At the same time, observation obtained by the ground-based telescopes may also be helpful for the research of such distant system. Thereby, the Laboratory of observational astrometry of Pulkovo Observatory of RAS made a decision to reprocess observations obtained during last decade. More than 350 positional observations of Pluto – Charon system were carried out with the mirror astrograph ZA-320M at Pulkovo and Maksutov telescope MTM-500M near Kislovodsk. These observations were processed by means of software system APEX-II developed in Pulkovo observatory and numerical simulation was performed to calculate the differences between positions of photocenter and barycenter of Pluto – Charon system.

**Introduction**

From 1999 till 2010 the Laboratory of observational astrometry of Pulkovo observatory was carrying out regular astrometrical observations of the Pluto – Charon system [1]. For the period from May 1999 till August 2006 the observations were performed with the mirror astrograph ZA-320 ($D$ = 320 mm, $F$ = 3200 mm) [2], equipped with CCD camera SBIG ST-6 (1999–2004) and CCD camera FLI IMG 1001E (2005–2006). The observations were carried out preferably near the meridian at the hour angle not exceeded ±1$^h$. The zenith angle of Pluto was about 70° due to the declination of Pluto and latitude of the observatory (Pulkovo, Saint-Petersburg, Russia). The expositions between 60 and 200 seconds were used to collect enough light. Total amount of positional observations with ZA-320M for this period of time is 164. Due to decrease of Pluto's declination, observations with ZA-320M had to be terminated, but they were resumed on June 2008 with the new telescope MTM-500M ($D$ = 500 mm, $F$ = 4100 mm) equipped with CCD camera SBIG STL 1001E and located on the lower latitude near the city of Kislovodsk (Northern Caucasus, Russia) [3]. For the period from June 2008 till August 2010 total amount of 248 observations was obtained.

**Observations processing**

For a current time ground-based astrometrical observations with CCD cameras allow calculating the position of the celestial bodies with an accuracy about 0″.01 [4]. In order to reach such accuracy it is necessary to calculate the correction, caused by the geometrical and photometrical features of observable objects. For the observations performed on the mid-sized telescopes such as ZA-320M and MTM-500M due to the limited resolution, it is impossible to separate the images of two components of Pluto – Charon system, therefore only the position of photocenter can be derived from obtained image. Due to mutual motion of Pluto and Charon, correction for the difference in the position of photocenter and barycenter should be calculated.

As it already was mentioned, total amount of 359 positional observations have been taken for the period from 1999 to 2010 with a help of ZA-320M and MTM-500M telescopes. Processing of obtained images was performed with a help of the software system APEX-II [5] developed in Pulkovo observatory. Depending on the observation's conditions, from 8 to 450 reference stars from recently published UCAC4 catalog [6] were chosen on each image. In order to take into account the difference between the positions of barycenter and photocenter of Pluto – Charon system, the following scheme was used in the current work. On the basis of the ephemerids presented on the NASA (JPL) site [7], the distance between centers of Pluto and Charon as well as positional angle of Charon on the geocentric celestial sphere were determined for a specific time moments. These calculations were implemented by means of software system EPOS8 developed in Pulkovo observatory [8]. Then obtained data were used to simulate the image of the Pluto – Charon system for time moment of each real observation by the means of the following method.

First of all, as it was already mentioned, it is not possible to obtain sharp image of minor planets with a help of ground-based observations due to limited angular resolution and atmospheric fluctuations. In order to take these factors to consideration, the distributions of the illumination of Pluto or Charon on the obtained images were approximated by Gaussian

$$l(x,y) = l_0 \exp\left(-\frac{x^2 + y^2}{\sigma^2}\right), \qquad (1)$$

where σ – the variable, which characterize the "blurriness" of the image, (x, y) – the position relatively to the center of the component from the component's center. The value of $l_0$ depends on the magnitudes of Pluto and Charon.

The barycenter of *n*-body system can be easily calculated from the relations between the components masses. The barycenter of the Pluto – Charon system lays on the straight line connecting components' centers on the distance form Pluto which can be calculated by means of the following formula:

$$R_1 = \frac{Rm_2}{m_1 + m_2}, \qquad (2)$$

where *R* – the distance between Pluto and Charon centers, $m_1$ and $m_2$ – masses of Pluto and Charon respectively [9].

According to geometrical, photometrical and physical characteristics of the system [10], the simulated frames for the different positional angles of Charon were obtained. Figure 1 shows real and simulated frames of the Pluto – Charon system. Photocenter of the system is corresponding to the centroid of the simulated frame, which can be calculated by means of raw moments as

$$x_c = \frac{\int_x \int_y x l(x,y) dx dy}{\int_x \int_y l(x,y) dx dy} \qquad y_c = \frac{\int_x \int_y y l(x,y) dx dy}{\int_x \int_y l(x,y) dx dy} \qquad (3)$$

The photocenter of the system for the specific time moment is marked as "×" on the Figure 1. At the same time with a help of (2), the position of the barycenter for this specific position can also be easily calculated (it is marked by "•" sign on Figure 1).

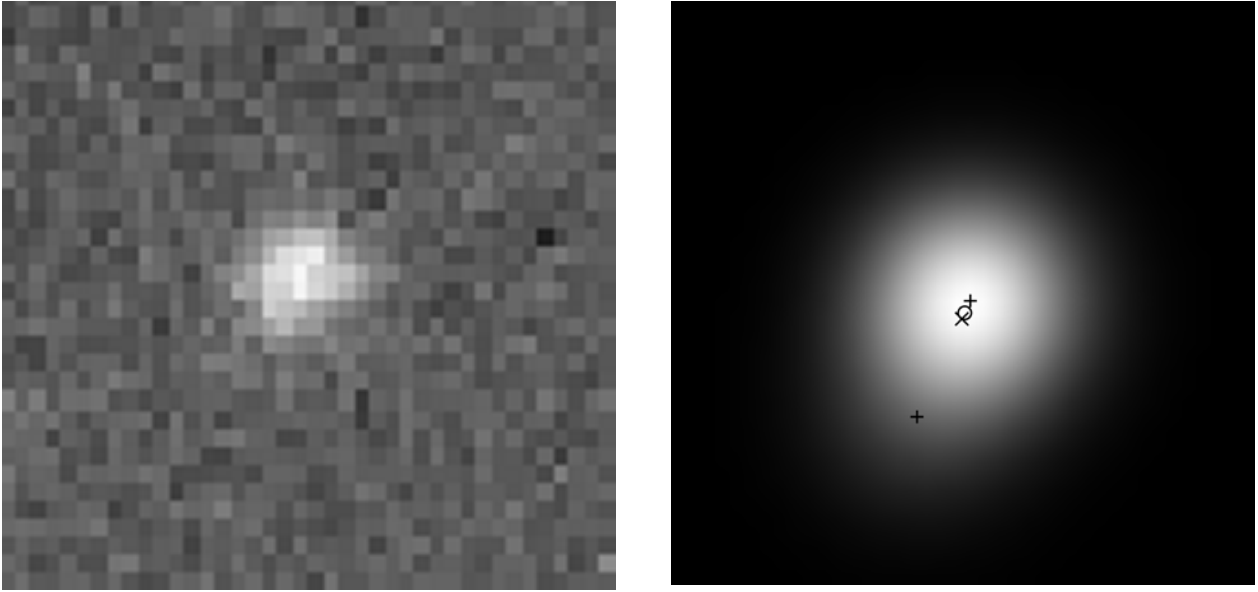

**Figure 1.** Real (left) and simulated (right) images of Pluto – Charon system for 27[th] March 2003 (not in the same scale). "+" symbols represent centers of the dwarf planets, "×" and "o" – photocenter and barycenter of the system respectively.

From these calculations, the difference between the positions of photo- and barycenters can be derived, which should be used during the processing of the real observations. It is obvious that due to the rotation of the system, the correctional values are depended on the time moment. Figure 2 represents the calculated corrections, decomposed on the correction for declination and right accession. This figure shows correction for the whole amount of observations, carried out for the period from 1999 till 2010 at ZA-320M and MTM-500M. The values of *X*-axis correspond to the positional angle for Charon, considering its rotation period of 6.38723 days [10].

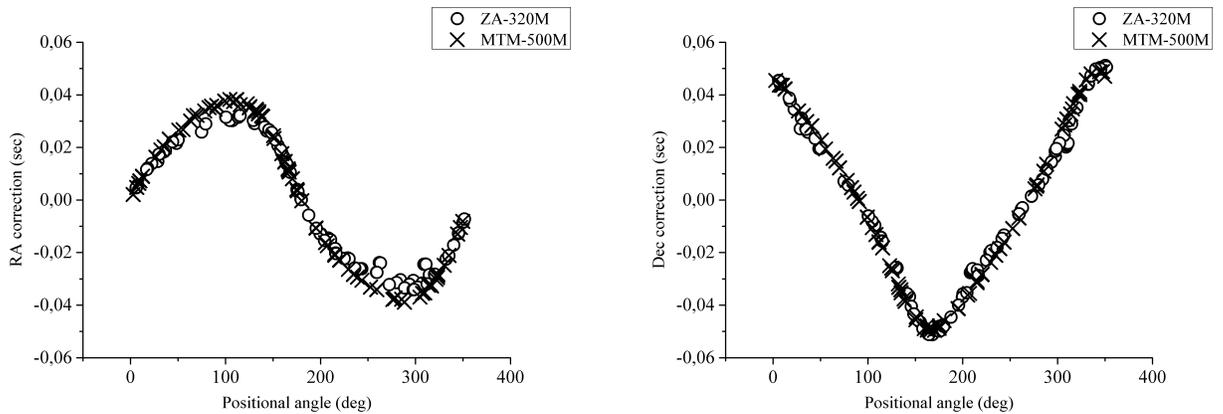

**Figure 2.** Corrections for the right accession (left) and declination (right) due to positional difference between photocenter and barycenter.

Obtained corrections can be summarized to estimate the O–C values. Table 1 shows average annual O–C values for the DE406 and DE431 ephemeris [11] with and without calculated corrections. Table 2 represents the similar data but average value calculated for one positional angle. Software system EPOS8 was used for the calculation of the ephemerides positions and O–C estimation.

**Table 1. Average annual O–C values of Pluto – Charon system, taking into account correction for barycenter's position ($N$ – number of images, "ZA" – ZA-320M, "MTM" – MTM-500M).**

| Observa-tional year | DE406 (O–C)$_\alpha$ s | DE406 (O–C)$_\delta$ " | DE406 bar(O–C)$_\alpha$ s | DE406 bar(O–C)$_\delta$ " | DE431 (O–C)$_\alpha$ s | DE431 (O–C)$_\delta$ " | $N$ | Tele-scope |
|---|---|---|---|---|---|---|---|---|
| 1999 | −0.068 ± 0.051 | 0.013 ± 0.029 | −0.088 ±.054 | 0.016 ± 0.026 | −0.036 ± 0.052 | 0.023 ± 0.029 | 13 | ZA |
| 2002 | −0.068 ± 0.034 | 0.271 ± 0.026 | −0.067 ± 0.035 | 0.272 ± 0.029 | −0.034 ± 0.034 | 0.277 ± 0.026 | 20 | ZA |
| 2003 | −0.063 ± 0.041 | 0.264 ± 0.044 | −0.062 ± 0.042 | 0.276 ± 0.044 | −0.025 ± 0.041 | 0.267 ± 0.044 | 27 | ZA |
| 2004 | −0.075 ± 0.029 | 0.319 ± 0.033 | −0.072 ± 0.030 | 0.328 ± 0.033 | −0.033 ± 0.029 | 0.317 ± 0.033 | 39 | ZA |
| 2005 | 0.044 ± 0.052 | 0.209 ± 0.040 | 0.025 ± 0.054 | 0.212 ± 0.037 | 0.094 ± 0.052 | 0.203 ± 0.040 | 12 | ZA |
| 2006 | −0.023 ± 0.082 | 0.093 ± 0.099 | −0.044 ± 0.082 | 0.096 ± 0.096 | 0.031 ± 0.082 | 0.083 ± 0.099 | 21 | ZA |
| 2008 | 0.076 ± 0.019 | 0.160 ± 0.010 | 0.079 ± 0.025 | 0.143 ± 0.020 | 0.154 ± 0.019 | 0.140 ± 0.010 | 7 | MTM |
| 2009 | −0.022 ± 0.009 | 0.133 ± 0.009 | −0.018 ± 0.009 | 0.144 ± 0.009 | 0.070 ± 0.009 | 0.107 ± 0.009 | 103 | MTM |
| 2010 | −0.093 ± 0.011 | 0.072 ± 0.010 | −0.094 ± 0.011 | 0.071 ± 0.010 | 0.017 ± 0.010 | 0.043 ± 0.010 | 117 | MTM |

**Table 2. Average O–C for one specific position of Pluto – Charon system, taking into account corrections for the barycenter's position (*N* – number of images, "ZA" – ZA-320M, "MTM" – MTM-500M).**

| Observational year | DE406 (O–C)$_\alpha$ s | DE406 (O–C)$_\delta$ " | DE406 bar(O–C)$_\alpha$ s | DE406 bar(O–C)$_\delta$ " | DE431 (O–C)$_\alpha$ s | DE431 (O–C)$_\delta$ " | N | Telescope |
|---|---|---|---|---|---|---|---|---|
| 1999 | 0.183 | 0.106 | 0.193 | 0.094 | 0.186 | 0.106 | 13 | ZA |
| 2002 | 0.152 | 0.115 | 0.156 | 0.129 | 0.153 | 0.116 | 20 | ZA |
| 2003 | 0.214 | 0.227 | 0.220 | 0.231 | 0.212 | 0.227 | 27 | ZA |
| 2004 | 0.182 | 0.206 | 0.187 | 0.208 | 0.182 | 0.207 | 39 | ZA |
| 2005 | 0.178 | 0.139 | 0.187 | 0.128 | 0.178 | 0.140 | 12 | ZA |
| 2006 | 0.374 | 0.452 | 0.375 | 0.438 | 0.375 | 0.452 | 21 | ZA |
| 2008 | 0.050 | 0.025 | 0.065 | 0.053 | 0.050 | 0.025 | 7 | MTM |
| 2009 | 0.091 | 0.095 | 0.095 | 0.096 | 0.089 | 0.094 | 103 | MTM |
| 2010 | 0.114 | 0.106 | 0.120 | 0.110 | 0.111 | 0.106 | 117 | MTM |

Topocentric astrometric coordinates of the Pluto – Charon system for J2000 epoch are presented in MPC format in the supplementary material section below.

**Conclusions**

For the period from 1999 till 2010 total amount of 359 observations of Pluto – Charon system was obtained. Consideration of the corrections for the difference between positions of barycenter and photocenter of Pluto – Charon system allows us to decrease O–C values for the DE406 system. These corrections can be derived from the relatively simple equations.

**References**

[1] A. V. Devyatkin, D. L. Gorshanov, V. V. Kouprianov, E. Y. Aleshkina, A. S. Bekhteva, G. D. Baturina, A. M. Ibragimov, I. A. Verestchagina, O. V. Krakosevich, and K. V. Barshevich, "Astrometrical observations of Uranus, Neptune and Pluto – Charon system with ZA-320M mirror astrograph in 2004-2006," *Izv. GAO RAN (Reports Pulkovo Obs.)*, vol. 218, pp. 87–93, 2006.

[2] A. V. Devyatkin, D. L. Gorshanov, V. V. Kouprianov, I. A. Vereshchagina, A. S. Bekhteva, and F. M. Ibragimov, "Astrometric and photometric observations of solar system bodies with Pulkovo Observatory's automatic mirror astrograph ZA-320M," *Sol. Syst. Res.*, vol. 43, no. 3, pp. 229–239, 2009.

[3] A. P. Kulish, A. V. Devyatkin, V. B. Rafalskiy, A. M. Ibragimov, V. V. Kouprianov, I. A. Verestchagina, and A. V. Shumakher, "Automation of MTM-500M telescope complex," *Izv. GAO RAN (Reports Pulkovo Obs.*, vol. 219, no. 1, pp. 192–218, 2009.

[4] P. Y. Bely, *The design and construction of large optical telescopes*. Heidelberg: Springer, 2003.

[5] A. V. Devyatkin, D. L. Gorshanov, V. V Kouprianov, and I. A. Verestchagina, "Apex I and Apex II Software Packages for the Reduction of Astronomical CCD Observations," vol. 44, no. 1, pp. 68–80, 2010.

[6] N. Zacharias, C. T. Finch, T. M. Girard, A. Henden, J. L. Bartlett, D. G. Monet, and M. I. Zacharias, "THE FOURTH US NAVAL OBSERVATORY CCD ASTROGRAPH CATALOG ( UCAC4 )," vol. 44, 2013.

[7] J. P. Laboratory, "JPL NASA Ephemerides." [Online]. Available: http://ssd.jpl.nasa.gov/?ephemerides.

[8] V. N. L and S. D. Tsekmeister, "The Use of the Epos Software Package for Research of the Solar System Objects 1," vol. 46, no. 2, pp. 190–192, 2012.

[9] D. J. Tholen, M. W. Buie, W. M. Grundy, and G. T. Elliott, "Masses of nix and hydra 1," no. 2006, pp. 777–784, 2008.

[10] W. H. Cheng, M. Hoi, and S. J. Peale, "Complete tidal evolution of Pluto – Charon," vol. 233, pp. 242–258, 2014.

[11] W. M. Folkner, J. G. Williams, D. H. Boggs, R. S. Park, and P. Kuchynka, *The Planetary and Lunar Ephemerides DE430 and DE431*. 2014.

**Supplementary material: Topocentric astrometric coordinates of Pluto-Charon system**

Topocentric astrometric coordinates of the Pluto – Charon system for J2000 epoch for observations on ZA-320M and MTM-500M are presented in the Tables A1 and A2 respectively (MPC format is used).

**Table A1. Topocentric astrometric coordinates of the Pluto – Charon system for J2000 epoch for observations on ZA-320M**

```
D4340          C1999 03 14.10400516 42 22.757-10 27 14.01         14.05      084
D4340          C1999 03 14.11459516 42 22.719-10 27 13.69         13.94      084
D4340          C1999 03 14.12125016 42 22.738-10 27 13.56         13.92      084
D4340          C1999 03 14.12688716 42 22.744-10 27 13.53         13.87      084
D4340          C1999 03 14.12971116 42 22.742-10 27 13.35         13.87      084
D4340          C1999 03 14.13814816 42 22.731-10 27 13.43         14.04      084
D4340          C1999 03 15.09319416 42 22.757-10 26 55.51         14.15      084
D4340          C1999 03 15.10605316 42 22.773-10 26 55.54         14.02      084
D4340          C1999 03 15.11678216 42 22.769-10 26 55.30         14.06      084
D4340          C1999 03 15.11959516 42 22.772-10 26 55.28         14.12      084
D4340          C1999 03 15.12240716 42 22.772-10 26 55.26         14.06      084
D4340          C1999 03 15.12523116 42 22.779-10 26 55.15         14.10      084
D4340          C1999 03 30.09182916 42 07.356-10 22 02.68         14.18      084
D4340          C2002 03 03.09751817 09 31.180-12 57 57.72         14.09      084
D4340          C2002 03 03.10501017 09 31.190-12 57 57.71         14.38      084
D4340          C2002 03 03.10880517 09 31.204-12 57 57.57         14.36      084
D4340          C2002 03 31.09281817 09 46.306-12 51 38.52         14.13      084
D4340          C2002 03 31.09796117 09 46.321-12 51 38.10         14.18      084
D4340          C2002 04 02.08487717 09 43.400-12 51 08.48         14.29      084
D4340          C2002 04 03.03498617 09 41.824-12 50 54.00         14.19      084
D4340          C2002 04 03.04098017 09 41.809-12 50 53.74         14.07      084
D4340          C2002 04 04.05113517 09 40.010-12 50 38.47         14.22      084
D4340          C2002 04 04.06248617 09 39.984-12 50 38.24         14.05      084
D4340          C2002 04 07.07308617 09 33.859-12 49 52.56         14.28      084
D4340          C2002 04 07.07537817 09 33.841-12 49 52.59         14.26      084
D4340          C2002 04 07.07736817 09 33.850-12 49 52.71         14.27      084
D4340          C2002 04 08.07994117 09 31.531-12 49 37.32         14.09      084
D4340          C2002 04 24.02943817 08 39.336-12 45 40.65         14.23      084
D4340          C2002 04 27.03342017 08 26.476-12 44 58.49         14.28      084
```

```
D4340          C2002 05 07.96057417 07 32.870-12 42 37.32          13.95      084
D4340          C2002 05 09.99149517 07 21.846-12 42 13.23          14.06      084
D4340          C2002 05 09.99548617 07 21.843-12 42 13.51          13.93      084
D4340          C2002 05 22.93267117 06 05.818-12 40 03.97          13.98      084
D4340          C2003 02 16.13856317 17 36.765-13 46 06.07          14.72      084
D4340          C2003 02 16.14094617 17 36.773-13 46 05.72          15.16      084
D4340          C2003 03 03.13621817 18 32.834-13 44 01.75          14.00      084
D4340          C2003 03 03.15030117 18 32.869-13 44 01.69          13.86      084
D4340          C2003 03 09.13342317 18 46.923-13 42 58.40          14.16      084
D4340          C2003 03 26.10128317 19 00.291-13 39 30.94          14.17      084
D4340          C2003 04 09.04771417 18 42.398-13 36 24.47          15.31      084
D4340          C2003 04 11.03409917 18 37.813-13 35 57.78          14.35      084
D4340          C2003 04 20.05617317 18 11.191-13 33 58.97          14.18      084
D4340          C2003 04 20.98624117 18 07.921-13 33 46.91          14.49      084
D4340          C2003 04 22.03585617 18 04.128-13 33 33.58          14.31      084
D4340          C2003 04 22.03901017 18 04.122-13 33 33.45          14.37      084
D4340          C2003 04 27.00208817 17 44.593-13 32 31.82          14.02      084
D4340          C2003 05 11.99640517 16 31.650-13 29 46.74          13.98      084
D4340          C2003 05 11.99866817 16 31.621-13 29 47.27          13.95      084
D4340          C2003 05 13.98409917 16 20.687-13 29 28.32          14.69      084
D4340          C2003 05 13.98632317 16 20.631-13 29 28.33          14.80      084
D4340          C2003 05 19.95624517 15 46.178-13 28 37.67          14.29      084
D4340          C2003 05 19.95883117 15 46.156-13 28 37.81          14.34      084
D4340          C2003 05 25.91327217 15 09.918-13 27 55.11          14.41      084
D4340          C2003 05 25.91599517 15 09.913-13 27 55.62          14.24      084
D4340          C2003 05 27.93844717 14 57.303-13 27 43.30          14.36      084
D4340          C2003 05 27.94524317 14 57.226-13 27 43.20          14.71      084
D4340          C2003 05 31.86268017 14 32.352-13 27 22.49          14.29      084
D4340          C2003 05 31.86657117 14 32.297-13 27 22.09          14.24      084
D4340          C2004 03 05.10755617 27 41.004-14 27 55.20          14.31      084
D4340          C2004 03 05.10987717 27 40.975-14 27 55.14          14.47      084
D4340          C2004 03 13.12046717 27 58.095-14 26 35.98          14.47      084
D4340          C2004 03 14.11900917 27 59.583-14 26 25.34          14.23      084
D4340          C2004 04 02.07545617 28 02.631-14 22 49.51          14.00      084
D4340          C2004 04 03.07411417 28 01.444-14 22 37.32          14.26      084
D4340          C2004 04 03.07755317 28 01.454-14 22 37.62          14.90      084
D4340          C2004 04 04.07774917 28 00.157-14 22 25.60          14.31      084
D4340          C2004 04 05.07431817 27 58.723-14 22 13.95          14.40      084
D4340          C2004 04 05.07746417 27 58.719-14 22 13.95          14.40      084
D4340          C2004 04 06.06757017 27 57.183-14 22 02.23          14.48      084
```

```
D4340        C2004 04 06.07005817 27 57.163-14 22 02.14          14.63       084
D4340        C2004 04 07.07623717 27 55.474-14 21 50.28          14.30       084
D4340        C2004 04 07.08338817 27 55.456-14 21 49.95          14.52       084
D4340        C2004 04 09.05964917 27 51.735-14 21 27.07          14.63       084
D4340        C2004 04 11.06206317 27 47.499-14 21 03.45          14.16       084
D4340        C2004 04 13.07374817 27 42.731-14 20 40.35          14.13       084
D4340        C2004 04 14.06291517 27 40.215-14 20 28.44          14.24       084
D4340        C2004 04 17.03896517 27 31.902-14 19 54.53          14.15       084
D4340        C2004 04 18.03737017 27 28.884-14 19 43.70          14.52       084
D4340        C2004 04 24.03053117 27 08.411-14 18 37.64          14.30       084
D4340        C2004 04 25.02937217 27 04.602-14 18 27.37                      084
D4340        C2004 04 25.03139217 27 04.635-14 18 26.96          14.00       084
D4340        C2004 04 26.02204617 27 00.767-14 18 16.41          14.15       084
D4340        C2004 04 26.04079817 27 00.695-14 18 16.04          14.23       084
D4340        C2004 04 27.02459617 26 56.720-14 18 05.42          14.23       084
D4340        C2004 05 08.97155517 26 01.677-14 16 12.81          14.25       084
D4340        C2004 05 08.97509817 26 01.635-14 16 12.73          14.58       084
D4340        C2004 05 09.96501917 25 56.541-14 16 04.74          14.25       084
D4340        C2004 05 12.96581117 25 40.595-14 15 40.37          14.50       084
D4340        C2004 05 12.96945017 25 40.582-14 15 40.34          14.20       084
D4340        C2004 05 13.95663617 25 35.200-14 15 32.85          14.61       084
D4340        C2004 05 16.97030817 25 18.317-14 15 11.09          13.84       084
D4340        C2004 05 21.91938317 24 49.411-14 14 39.92          15.04       084
D4340        C2004 05 21.94500917 24 49.254-14 14 39.39          14.12       084
D4340        C2004 05 23.94053417 24 37.252-14 14 28.35          14.33       084
D4340        C2004 05 23.94258917 24 37.237-14 14 28.34          14.31       084
D4340        C2004 09 10.75316317 17 07.363-14 33 51.51          14.64       084
D4340        C2004 09 10.75468017 17 07.350-14 33 51.50          14.52       084
D4340        C2005 04 02.07680617 37 10.088-15 06 03.03          14.33       084
D4340        C2005 04 02.07831417 37 10.076-15 06 02.96          14.58       084
D4340        C2005 04 02.08077717 37 10.115-15 06 02.88          14.63       084
D4340        C2005 04 02.08228517 37 10.074-15 06 02.95          14.57       084
D4340        C2005 04 15.05555017 36 48.894-15 03 51.63          14.62       084
D4340        C2005 04 25.02270017 36 18.747-15 02 17.86          14.56       084
D4340        C2005 04 25.02373417 36 18.729-15 02 17.69          14.69       084
D4340        C2005 05 09.97527617 35 14.001-15 00 18.93          14.72       084
D4340        C2005 05 09.97631217 35 13.989-15 00 19.21          14.30       084
D4340        C2005 05 15.96457117 34 42.590-14 59 41.77          15.16       084
D4340        C2005 05 21.94206317 34 08.776-14 59 11.93          14.22       084
D4340        C2005 05 22.95458917 34 02.863-14 59 07.17          14.61       084
```

```
D4340          C2006 07 04.87920717 38 50.157-15 43 19.94          14.28          084
D4340          C2006 07 04.88175217 38 50.151-15 43 19.99          14.66          084
D4340          C2006 07 04.88206217 38 50.195-15 43 20.17          14.83          084
D4340          C2006 07 04.88269717 38 50.161-15 43 20.10          15.48          084
D4340          C2006 07 04.88300817 38 50.172-15 43 20.61          14.84          084
D4340          C2006 07 04.88364717 38 50.178-15 43 20.00          14.80          084
D4340          C2006 07 04.88396217 38 50.166-15 43 20.45          14.82          084
D4340          C2006 07 04.88427417 38 50.181-15 43 20.10          14.74          084
D4340          C2006 07 04.88458717 38 50.194-15 43 20.05          15.30          084
D4340          C2006 07 05.87756517 38 44.132-15 43 27.11          15.05          084
D4340          C2006 07 05.87787817 38 44.154-15 43 26.99          15.25          084
D4340          C2006 07 05.87861317 38 44.084-15 43 27.68                         084
D4340          C2006 07 05.87942917 38 44.128-15 43 27.24          13.97          084
D4340          C2006 07 05.87974317 38 44.110-15 43 26.78          14.80          084
D4340          C2006 07 05.88005917 38 44.052-15 43 26.94          14.78          084
D4340          C2006 07 05.88037017 38 44.062-15 43 27.35          13.96          084
D4340          C2006 07 05.89737517 38 43.977-15 43 28.04          15.25          084
D4340          C2006 07 05.89804917 38 43.963-15 43 27.32          14.60          084
D4340          C2006 07 05.89838017 38 44.010-15 43 27.49          14.63          084
D4340          C2006 07 06.89832717 38 37.936-15 43 34.73          14.55          084
D4340          C2006 07 06.89869517 38 37.973-15 43 36.13          14.68          084
```

**Table A2. Topocentric astrometric coordinates of the Pluto – Charon system for J2000 epoch for observations on MTM-500M**

```
D4340          C2008 07 01.96708917 57 34.195-17 03 45.52          14.46          C20
D4340          C2008 07 03.93978717 57 21.757-17 04 02.20          15.30          C20
D4340          C2008 07 03.94163317 57 21.746-17 04 02.19          14.93          C20
D4340          C2008 07 04.81565417 57 16.294-17 04 09.82          15.12          C20
D4340          C2008 07 04.81750017 57 16.282-17 04 09.88          15.06          C20
D4340          C2008 07 05.81904517 57 10.047-17 04 18.74          14.75          C20
D4340          C2008 07 05.82088617 57 10.037-17 04 18.75          14.79          C20
D4340          C2009 05 01.97955118 12 25.289-17 37 32.36          14.32          C20
D4340          C2009 05 01.98103618 12 25.286-17 37 32.26          14.32          C20
D4340          C2009 05 14.94888018 11 32.725-17 37 22.38          14.01          C20
D4340          C2009 05 14.95036618 11 32.718-17 37 22.40          14.12          C20
D4340          C2009 05 16.00761118 11 27.734-17 37 22.85          14.05          C20
D4340          C2009 05 16.99921218 11 22.973-17 37 23.37          14.00          C20
D4340          C2009 05 17.00219318 11 22.959-17 37 23.38          14.03          C20
```

```
D4340          C2009 05 17.00367918 11 22.951-17 37 23.37          14.00           C20
D4340          C2009 05 19.02063618 11 13.009-17 37 25.00          14.10           C20
D4340          C2009 05 23.94939118 10 47.382-17 37 32.05          14.07           C20
D4340          C2009 05 23.95087418 10 47.380-17 37 32.10          14.05           C20
D4340          C2009 05 29.96862618 10 13.801-17 37 46.68          13.98           C20
D4340          C2009 05 29.97011218 10 13.792-17 37 46.67          14.03           C20
D4340          C2009 05 30.89136218 10 08.472-17 37 49.64          14.07           C20
D4340          C2009 06 01.88415618 09 56.775-17 37 56.05          14.03           C20
D4340          C2009 06 01.88564318 09 56.761-17 37 56.02          14.05           C20
D4340          C2009 06 02.87801718 09 50.864-17 37 59.65          14.06           C20
D4340          C2009 06 02.87950018 09 50.856-17 37 59.63          14.01           C20
D4340          C2009 06 03.87697418 09 44.885-17 38 03.43          14.01           C20
D4340          C2009 06 03.87845318 09 44.876-17 38 03.48          14.00           C20
D4340          C2009 06 05.94425518 09 32.353-17 38 11.86          13.96           C20
D4340          C2009 06 05.94573918 09 32.346-17 38 11.90          13.96           C20
D4340          C2009 06 10.85947218 09 01.921-17 38 34.87                          C20
D4340          C2009 06 10.86096918 09 01.913-17 38 34.96                          C20
D4340          C2009 06 11.85768518 08 55.643-17 38 40.13                          C20
D4340          C2009 06 11.85916518 08 55.634-17 38 40.17                          C20
D4340          C2009 06 21.88063718 07 51.451-17 39 42.90                          C20
D4340          C2009 06 21.88213118 07 51.441-17 39 42.93                          C20
D4340          C2009 06 22.87513218 07 45.024-17 39 50.15                          C20
D4340          C2009 06 22.87661818 07 45.015-17 39 50.23                          C20
D4340          C2009 06 22.93351918 07 44.641-17 39 50.56                          C20
D4340          C2009 06 22.93500318 07 44.631-17 39 50.63                          C20
D4340          C2009 06 23.82064318 07 38.908-17 39 57.16                          C20
D4340          C2009 06 23.82212518 07 38.898-17 39 57.17                          C20
D4340          C2009 06 23.93052918 07 38.194-17 39 58.03                          C20
D4340          C2009 06 23.93201418 07 38.185-17 39 58.07                          C20
D4340          C2009 06 24.82071818 07 32.440-17 40 04.72                          C20
D4340          C2009 06 24.82219918 07 32.430-17 40 04.77                          C20
D4340          C2009 06 24.93894718 07 31.674-17 40 05.72                          C20
D4340          C2009 06 24.94043318 07 31.662-17 40 05.64                          C20
D4340          C2009 06 29.93232418 06 59.515-17 40 46.52                          C20
D4340          C2009 06 29.93381818 06 59.506-17 40 46.55                          C20
D4340          C2009 07 01.83407718 06 47.369-17 41 03.21          13.91           C20
D4340          C2009 07 01.94051518 06 46.681-17 41 04.24          13.98           C20
D4340          C2009 07 01.94199718 06 46.673-17 41 04.22          14.02           C20
D4340          C2009 07 02.90046718 06 40.582-17 41 12.91          14.08           C20
D4340          C2009 07 02.90195018 06 40.575-17 41 12.88          14.07           C20
```

```
D4340          C2009 07 03.87201218 06 34.430-17 41 21.83          14.17          C20
D4340          C2009 07 04.91729118 06 27.836-17 41 31.45                         C20
D4340          C2009 07 04.91877418 06 27.827-17 41 31.70          14.19          C20
D4340          C2009 07 07.96079718 06 08.821-17 42 01.15                         C20
D4340          C2009 07 07.96227918 06 08.813-17 42 01.14                         C20
D4340          C2009 07 08.78265318 06 03.732-17 42 09.33          14.13          C20
D4340          C2009 07 08.78413518 06 03.725-17 42 09.46          14.24          C20
D4340          C2009 07 11.82765518 05 45.135-17 42 41.01          14.20          C20
D4340          C2009 07 12.87550618 05 38.818-17 42 52.19          13.55          C20
D4340          C2009 07 12.87663918 05 38.814-17 42 52.21          14.01          C20
D4340          C2009 07 13.86662718 05 32.907-17 43 02.82          14.63          C20
D4340          C2009 07 13.87022118 05 32.873-17 43 03.00          14.23          C20
D4340          C2009 07 13.87170818 05 32.863-17 43 03.06          13.96          C20
D4340          C2009 08 12.81223118 03 05.672-17 49 33.26                         C20
D4340          C2009 08 12.81351718 03 05.670-17 49 33.23          14.19          C20
D4340          C2009 08 23.79021218 02 32.769-17 52 21.37          14.84          C20
D4340          C2009 08 23.79063718 02 32.772-17 52 21.36          14.87          C20
D4340          C2009 08 23.79106218 02 32.770-17 52 21.32          14.79          C20
D4340          C2009 08 23.79148618 02 32.767-17 52 21.42          14.91          C20
D4340          C2009 08 25.72243518 02 28.408-17 52 51.93          14.79          C20
D4340          C2009 08 25.72286018 02 28.410-17 52 51.84          14.84          C20
D4340          C2009 08 25.72328018 02 28.406-17 52 51.98          14.74          C20
D4340          C2009 08 25.72370318 02 28.407-17 52 51.94          14.85          C20
D4340          C2009 08 25.78414118 02 28.290-17 52 52.97          14.96          C20
D4340          C2009 08 25.78456818 02 28.293-17 52 52.95          15.10          C20
D4340          C2009 08 25.78499018 02 28.290-17 52 53.05          15.01          C20
D4340          C2009 08 25.78541418 02 28.293-17 52 52.88          15.10          C20
D4340          C2009 08 29.75048118 02 20.800-17 53 56.56          15.09          C20
D4340          C2009 08 29.75095018 02 20.801-17 53 56.48          15.09          C20
D4340          C2009 08 29.75141818 02 20.801-17 53 56.45          15.29          C20
D4340          C2009 08 29.75188618 02 20.800-17 53 56.56          14.99          C20
D4340          C2009 08 30.71969218 02 19.266-17 54 11.86          14.98          C20
D4340          C2009 08 30.72011718 02 19.265-17 54 11.89          14.68          C20
D4340          C2009 08 30.72054218 02 19.265-17 54 11.86          14.73          C20
D4340          C2009 08 30.72096718 02 19.264-17 54 11.91          14.70          C20
D4340          C2009 08 31.77572518 02 17.737-17 54 28.99          15.24          C20
D4340          C2009 08 31.77614918 02 17.733-17 54 29.14          15.36          C20
D4340          C2009 08 31.77657418 02 17.741-17 54 29.10          15.21          C20
D4340          C2009 08 31.77700318 02 17.730-17 54 29.14          15.39          C20
D4340          C2009 09 12.74446818 02 10.225-17 57 45.42          14.61          C20
```

```
D4340         C2009 09 12.74489318 02 10.229-17 57 45.52          14.72         C20
D4340         C2009 09 13.73543318 02 10.439-17 58 01.62          14.94         C20
D4340         C2009 09 13.73586118 02 10.448-17 58 01.49          14.98         C20
D4340         C2009 09 13.73628818 02 10.446-17 58 01.58          15.17         C20
D4340         C2009 09 14.75781418 02 10.806-17 58 18.35          15.08         C20
D4340         C2009 09 14.75823918 02 10.801-17 58 18.37          14.89         C20
D4340         C2009 09 14.75866818 02 10.798-17 58 18.46          14.85         C20
D4340         C2009 09 14.75909918 02 10.804-17 58 18.39          15.01         C20
D4340         C2009 09 18.76055018 02 13.520-17 59 24.57          15.17         C20
D4340         C2009 09 18.76097218 02 13.525-17 59 24.46          15.01         C20
D4340         C2009 09 18.76139518 02 13.522-17 59 24.56          14.96         C20
D4340         C2009 09 18.76181818 02 13.521-17 59 24.55          14.97         C20
D4340         C2009 10 12.67865818 03 13.738-18 05 48.92          14.61         C20
D4340         C2009 10 12.67908318 03 13.738-18 05 48.95          14.56         C20
D4340         C2009 10 12.67950618 03 13.743-18 05 48.94          14.55         C20
D4340         C2009 10 12.67992818 03 13.746-18 05 48.93          14.59         C20
D4340         C2010 02 11.13153518 18 54.291-18 16 56.13          14.65         C20
D4340         C2010 02 11.13196118 18 54.281-18 16 56.35          14.62         C20
D4340         C2010 02 11.13238718 18 54.291-18 16 56.38          14.56         C20
D4340         C2010 02 11.13281318 18 54.299-18 16 56.33          14.58         C20
D4340         C2010 02 24.09525418 20 12.604-18 15 53.37          15.24         C20
D4340         C2010 02 24.09569118 20 12.605-18 15 53.35          15.22         C20
D4340         C2010 02 24.09616418 20 12.611-18 15 53.42          15.24         C20
D4340         C2010 02 24.09663718 20 12.608-18 15 53.41          15.35         C20
D4340         C2010 02 26.08906218 20 23.022-18 15 42.88          15.09         C20
D4340         C2010 03 02.10866818 20 42.631-18 15 21.54          14.97         C20
D4340         C2010 03 02.10909418 20 42.631-18 15 21.46          15.04         C20
D4340         C2010 03 02.10952218 20 42.633-18 15 21.43          14.90         C20
D4340         C2010 03 02.10995118 20 42.634-18 15 21.46          14.96         C20
D4340         C2010 03 03.12069118 20 47.270-18 15 16.06          14.66         C20
D4340         C2010 03 03.12111918 20 47.280-18 15 16.17          14.43         C20
D4340         C2010 03 03.12154518 20 47.275-18 15 16.02          14.47         C20
D4340         C2010 03 03.12197118 20 47.287-18 15 16.13          14.37         C20
D4340         C2010 03 24.02391218 21 54.135-18 13 25.92          14.95         C20
D4340         C2010 03 24.02434018 21 54.137-18 13 25.94          14.81         C20
D4340         C2010 03 24.02476718 21 54.135-18 13 25.94          14.93         C20
D4340         C2010 03 24.02519318 21 54.136-18 13 25.91          14.87         C20
D4340         C2010 03 25.01794918 21 55.894-18 13 21.03          14.96         C20
D4340         C2010 03 25.01837518 21 55.889-18 13 21.04          14.90         C20
D4340         C2010 03 25.01880318 21 55.893-18 13 21.08          15.02         C20
```

```
D4340          C2010 03 25.01923218 21 55.895-18 13 20.87          15.07        C20
D4340          C2010 03 27.08508218 21 59.110-18 13 11.36          14.92        C20
D4340          C2010 03 27.08550518 21 59.110-18 13 11.34          14.50        C20
D4340          C2010 03 27.08592918 21 59.107-18 13 11.27          14.91        C20
D4340          C2010 03 27.08635918 21 59.110-18 13 11.34          14.68        C20
D4340          C2010 03 28.03261318 22 00.403-18 13 06.96          15.01        C20
D4340          C2010 03 28.03304318 22 00.412-18 13 07.02          15.03        C20
D4340          C2010 03 28.03347018 22 00.402-18 13 06.89          15.01        C20
D4340          C2010 03 28.03390118 22 00.398-18 13 06.81          14.98        C20
D4340          C2010 03 29.01250518 22 01.606-18 13 02.48          15.07        C20
D4340          C2010 03 29.01293218 22 01.605-18 13 02.48          15.05        C20
D4340          C2010 03 29.01336318 22 01.609-18 13 02.39          15.05        C20
D4340          C2010 03 29.01379218 22 01.609-18 13 02.47          15.05        C20
D4340          C2010 04 01.07774718 22 04.546-18 12 49.05          14.40        C20
D4340          C2010 04 01.07817618 22 04.546-18 12 49.10          14.51        C20
D4340          C2010 04 01.07860618 22 04.557-18 12 49.07          14.42        C20
D4340          C2010 04 01.07903618 22 04.550-18 12 48.98          14.28        C20
D4340          C2010 04 07.06970618 22 06.702-18 12 24.92          14.78        C20
D4340          C2010 04 07.07013218 22 06.702-18 12 25.10          14.80        C20
D4340          C2010 04 07.07056018 22 06.702-18 12 24.96          14.85        C20
D4340          C2010 04 07.07099318 22 06.703-18 12 25.03          14.85        C20
D4340          C2010 04 09.05984818 22 06.359-18 12 17.89          14.77        C20
D4340          C2010 04 09.06027518 22 06.359-18 12 17.77          14.95        C20
D4340          C2010 04 09.06070518 22 06.363-18 12 17.78          15.00        C20
D4340          C2010 04 09.06120918 22 06.357-18 12 17.78          14.96        C20
D4340          C2010 04 16.03299318 22 01.061-18 11 56.00          14.30        C20
D4340          C2010 04 16.03342118 22 01.062-18 11 56.00          15.03        C20
D4340          C2010 04 16.03385218 22 01.061-18 11 56.08          15.01        C20
D4340          C2010 04 16.03428318 22 01.062-18 11 56.09          14.46        C20
D4340          C2010 04 18.02980618 21 58.388-18 11 51.27          14.12        C20
D4340          C2010 04 18.03023718 21 58.387-18 11 51.36          14.14        C20
D4340          C2010 04 18.03067018 21 58.384-18 11 51.35          14.12        C20
D4340          C2010 04 18.03110218 21 58.388-18 11 51.30          14.08        C20
D4340          C2010 04 19.94907618 21 55.344-18 11 46.56          14.75        C20
D4340          C2010 04 19.94950518 21 55.345-18 11 46.53          14.75        C20
D4340          C2010 04 19.94993218 21 55.338-18 11 46.54          14.76        C20
D4340          C2010 04 19.95036318 21 55.347-18 11 46.66          14.81        C20
D4340          C2010 04 28.92195418 21 35.156-18 11 32.82          14.58        C20
D4340          C2010 04 28.92238118 21 35.152-18 11 32.79          14.62        C20
D4340          C2010 04 28.92281318 21 35.155-18 11 32.80          14.66        C20
```

```
D4340         C2010 04 29.985264 18 21 32.132 -18 11 32.11          14.84        C20
D4340         C2010 04 29.985691 18 21 32.132 -18 11 31.99          14.76        C20
D4340         C2010 04 29.986121 18 21 32.132 -18 11 32.01          14.81        C20
D4340         C2010 04 29.986549 18 21 32.131 -18 11 32.01          14.75        C20
D4340         C2010 07 26.928236 18 13 26.686 -18 21 07.49          14.00        C20
D4340         C2010 07 26.928658 18 13 26.679 -18 21 07.56          14.07        C20
D4340         C2010 07 28.905905 18 13 16.344 -18 21 33.53          14.53        C20
D4340         C2010 07 28.906349 18 13 16.354 -18 21 33.36          14.53        C20
D4340         C2010 07 28.906794 18 13 16.336 -18 21 33.37          14.46        C20
D4340         C2010 07 28.907239 18 13 16.340 -18 21 33.38          14.55        C20
D4340         C2010 07 29.866131 18 13 11.434 -18 21 46.22          14.51        C20
D4340         C2010 07 29.866558 18 13 11.428 -18 21 46.33          14.52        C20
D4340         C2010 07 29.867412 18 13 11.428 -18 21 46.21          14.51        C20
D4340         C2010 07 31.877776 18 13 01.371 -18 22 13.21                       C20
D4340         C2010 07 31.878201 18 13 01.367 -18 22 13.20                       C20
D4340         C2010 07 31.878626 18 13 01.362 -18 22 13.25                       C20
D4340         C2010 07 31.879051 18 13 01.362 -18 22 13.24                       C20
D4340         C2010 08 01.837603 18 12 56.684 -18 22 26.46          14.37        C20
D4340         C2010 08 01.838026 18 12 56.690 -18 22 26.22          14.48        C20
D4340         C2010 08 01.838452 18 12 56.683 -18 22 26.26          14.44        C20
D4340         C2010 08 01.838875 18 12 56.683 -18 22 26.21          14.43        C20
D4340         C2010 08 05.803734 18 12 38.121 -18 23 20.95                       C20
D4340         C2010 08 05.804167 18 12 38.148 -18 23 20.98                       C20
D4340         C2010 08 05.805035 18 12 38.126 -18 23 21.22                       C20
D4340         C2010 08 06.726266 18 12 34.002 -18 23 33.79          14.38        C20
D4340         C2010 08 06.726687 18 12 33.998 -18 23 33.80          14.17        C20
D4340         C2010 08 06.727109 18 12 33.999 -18 23 33.83          14.31        C20
D4340         C2010 08 06.727529 18 12 33.996 -18 23 33.79          14.21        C20
D4340         C2010 08 07.769079 18 12 29.430 -18 23 48.49          14.24        C20
D4340         C2010 08 07.769935 18 12 29.423 -18 23 48.73                       C20
D4340         C2010 08 09.738717 18 12 21.081 -18 24 16.53          14.11        C20
D4340         C2010 08 09.739138 18 12 21.078 -18 24 16.50          14.05        C20
D4340         C2010 08 09.739561 18 12 21.077 -18 24 16.45          14.10        C20
D4340         C2010 08 10.809664 18 12 16.712 -18 24 31.64          14.26        C20
D4340         C2010 08 10.810085 18 12 16.712 -18 24 31.71          14.22        C20
D4340         C2010 08 10.810507 18 12 16.709 -18 24 31.71          14.59        C20
D4340         C2010 08 10.810930 18 12 16.708 -18 24 31.85          14.63        C20
D4340         C2010 08 23.737019 18 11 33.295 -18 27 42.27          14.65        C20
D4340         C2010 08 23.737451 18 11 33.289 -18 27 42.26          14.56        C20
D4340         C2010 08 23.737909 18 11 33.288 -18 27 42.28          14.63        C20
```

```
D4340         C2010 08 23.73834418 11 33.287-18 27 42.26          14.68        C20
D4340         C2010 08 24.83812518 11 30.442-18 27 58.88          14.66        C20
D4340         C2010 08 24.83855418 11 30.450-18 27 58.62          14.57        C20
D4340         C2010 08 24.83898118 11 30.438-18 27 58.85          14.63        C20
D4340         C2010 08 24.83940918 11 30.425-18 27 58.63          14.71        C20
D4340         C2010 08 29.72653318 11 19.533-18 29 12.95          14.37        C20
D4340         C2010 08 29.72696218 11 19.532-18 29 12.98          14.45        C20
D4340         C2010 08 29.72739118 11 19.534-18 29 13.03          14.51        C20
D4340         C2010 08 29.72782218 11 19.532-18 29 13.01          14.54        C20
D4340         C2010 08 30.74196618 11 17.620-18 29 28.39          14.51        C20
D4340         C2010 08 30.74238818 11 17.626-18 29 28.53          14.59        C20
D4340         C2010 08 30.74281318 11 17.626-18 29 28.53          14.55        C20
D4340         C2010 08 30.74323718 11 17.620-18 29 28.28          14.58        C20
```